\documentclass[%
 reprint,
superscriptaddress,
nofootinbib,
 amsmath,amssymb,
 aps,
 pra,
]{revtex4-1}
\usepackage{verbatim}
\usepackage{mathtools}
\usepackage{amsthm}
\usepackage{physics}
\usepackage{graphicx}
\usepackage{dcolumn}
\usepackage{bm}
\usepackage{xfrac}
\graphicspath{ {./images/} }

\begin{document}


\title{Analysis of quantum correlations within the ground state of a three level Lipkin model}

\author{Javier Faba}

\affiliation{%
Center for Computational Simulation, Universidad Polit\'ecnica de Madrid, Campus Montegancedo, 28660 Boadilla del Monte, Madrid, Spain
}%

\affiliation{%
Departamento de F\'isica Te\'orica and CIAFF, Universidad Aut\'onoma de Madrid, E-28049 Madrid, Spain 
}%

\author{Vicente Mart\'\i n}

\affiliation{%
Center for Computational Simulation, Universidad Polit\'ecnica de Madrid, Campus Montegancedo, 28660 Boadilla del Monte, Madrid, Spain
}%

\author{Luis Robledo}
\affiliation{%
Departamento de F\'isica Te\'orica and CIAFF, Universidad Aut\'onoma de Madrid, E-28049 Madrid, Spain 
}%

\affiliation{%
Center for Computational Simulation, Universidad Polit\'ecnica de Madrid, Campus Montegancedo, 28660 Boadilla del Monte, Madrid, Spain
}%

\date{\today}

\begin{abstract}

The performance of beyond mean field  methods in solving the quantum many body problem for fermions is usually characterized by the correlation energy measured with respect to the underlying mean field value. In this paper we address the issue of characterizing the amount of correlations associated to different approximations from a quantum information perspective. With this goal in mind, we analyze the traditional Hartree-Fock (HF) method with spontaneous symmetry breaking, the HF with symmetry restoration and the generator coordinate method (GCM) in a exactly solvable fermion model known as the three-level Lipkin model. To characterize correlations including entanglement and beyond we use the quantum discord between different partition orbitals. We find that for physically motivated partitions, the quantum discord of the exact ground state is reasonably well reproduced by the different approximations. However, other partitions create ``fake quantum correlations"  in order to capture quantum correlations corresponding to partitions for which the Hartree-Fock solution fails. Those are removed and redistributed through a symmetry restoration process.

\end{abstract}

\pacs{Valid PACS appear here}
\maketitle


\section{\label{sec:introduction}Introduction}

In 1814, Pierre-Simon Laplace stated in his work `A Philosophical Essay on Probabilities' that if an intelligence (which he called demon) had the ability to know the position and momentum of each particle in the universe at a given time, then he will be able to predict the trajectory of every body in the universe and nothing would be uncertain for him \cite{laplace}. In quantum mechanics Laplace's demon would have to deal with the absence of the concept of `trajectory', and the quantum mechanical formulation of Laplace's thought experiment would be a demon who knew the wave function of the entire universe at a given time. However, apart from the probabilistic interpretation of quantum mechanics, the knowledge of the whole doesn't imply the knowledge of its parts in a quantum system. This is the essence of entanglement between different partitions of the system. To measure entanglement quantities such as the entanglement entropy are introduced. It measures the entanglement between two partitions of the system by looking at  the amount of information lost in one partition when all the information about the other partition is forgotten. In this way, for a general state, the Laplace's demon would be uncertain about the partitions despite knowing the wave function of the whole system. Hence, computing this kind of purely quantum correlations (or `uncertainties') between the system's constituents is useful in order to analyze the separability, correlations, and therefore, the quantum structure of a system.

Entanglement is an important ingredient in the understanding of the performance of different approximations to the quantum many body problem. The size of the Hilbert space of a many body system scales exponentially with the number of particles in the system, preventing, for instance the complete understanding of even very simple molecules. Fortunately, the ground state and low lying states of such systems are supposed to ``live" in a corner of the Hilbert space accessible to variational techniques. In fermion systems, the use of Slater determinants as variational space leads to the Hartree-Fock (HF) approximation  \cite{ring_schuck}, that turns out to be a very good starting point for subsequent refinements \footnote{We can improve the Slater determinant ansatz using quasiparticle vacuum (Hartree-Fock-Bogoliubov or HFB) states instead. Those are the more general non-interacting particle states. However, for the three level Lipkin model, the results obtained from HFB states are equivalent to the HF ones, so we won't consider them.}. One of the defining properties of HF is the spontaneous symmetry breaking mechanism, where the single particle orbitals characterizing the HF Slater determinants break the symmetries of the Hamiltonian \cite{ring_schuck,symmetry_restoration_luis}. This is a simple mechanism used to consider additional correlations while maintaining the simplicity of the mean field wave functions. However, proper quantum numbers of the symmetries of the Hamiltonian have to be restored and the adequate framework is that of using projection operators leading to a linear combination of the wave functions obtained by applying the symmetry operator to the symmetry breaking mean field wave functions. The method, denoted by the name of Projected Hartree-Fock (PHF), works because the action of the symmetry operator on the symmetry breaking mean field is again a symmetry breaking mean field wave function.

On the other hand, we can extend the idea of using as an ansatz  linear combinations of not necessarily orthogonal trial states (such as Slater determinants), by postulating a general wave function
\begin{equation*}
    \ket{\psi} = \int da f(a)\ket{\phi(a)}
\end{equation*}
with the $f(a)$ amplitudes determined by the variational principle. This method is called the generator coordinate method (GCM) and is being used in many branches of quantum many body physics under different names. The success of the GCM ansatz is intimately associated to the adequateness of the the trial states $\ket{\phi(a)}$ for the physics one wants to describe. With a proper choice, the GCM ground state and/or excited states spectrum can be very close or identical to the exact solution.

Each of the above mentioned approximation methods HF, PHF and GCM incorporate different kinds of correlations into the description of the system. As the three of them are variational, a possibility to measure correlation is the use of the correlation energy, which is defined as the difference between the reference state energy (which can be, for example, the exact ground state) and the mean field one. In this way, the correlation energy for the mean field state will be always zero by definition, and maximum for the exact ground state. If we recover the symmetries of the Hamiltonian through a symmetry restoration process or performing a GCM, the correlation energy of the resulting approximate state will acquire a value between zero and the correlation energy of the exact ground state. In this way, a symmetry projection or the GCM will recover a fraction of the total correlations within a system. Another possibility is to use the quantum information perspective, where correlations are understood in a very different way: they are related with the separability of a system and the information/uncertainty gained when acting on a part of it. Given a $A|B$ bi-partition of a system, quantities such as the entanglement entropy, measure the uncertainty (i.e, non purity) of the $A (B)$ subsystem through the partial trace operation over the $B(A)$ part \cite{horodecki,adesso,gigena_overall_entropy,gigena_manybody_entanglement}. On the other hand, quantities such as the quantum discord \cite{quantum_discord,henderson} measure the correlations through a projective quantum measurement within the part $A(B)$. With this, a question arises: Is it possible to find a link between the correlations related to the spontaneous symmetry breaking mechanism within the mean-field/symmetry restoration context  (the ones that are related to the interaction of the constituents) and the correlations described by quantum information theory (the ones that are related to the uncertainty of a subsystem)?


In this work we will analyze the structure of the quantum correlations of an analytically solvable many-body system given by the three level Lipkin model by analyzing the exact, HF, PHF and GCM ground states. We will compute the quantum discord for multiple physically motivated partitions and we will link the results with the spontaneous symmetry breaking mechanism. The use of the three level Lipkin model is motivated by the fact that it is an exactly solvable but non-trivial model with two phase transitions due to the parity symmetry breaking \cite{HOLZWARTH,three_lipkin_shou,bertsch}. In Sec. \ref{sec:theory} we will introduce the theoretical concepts that will be discussed. In Sec. \ref{sec:3_level_lipkin} we will explain the three level Lipkin model and the results obtained. Finally, Sec. \ref{sec:conclusions} is a summary of the conclusions obtained in this work.


\section{\label{sec:theory}Theoretical background}


There are several ways to describe correlations under a quantum information context in a fermion many-body system \cite{fermions_cirac,yoshiko,gigena_overall_entropy,friis_relativity,fermionic_partial_trace,gigena_manybody_entanglement}. Due to the antisymmetrization principle, the Fock space is not defined as a tensor product of Hilbert spaces and therefore we cannot use the von Neumann entropy of the particle-reduced density matrix as a measure of entanglement: the components (particles) of a fermion system are indistinguishable, so it wouldn't make physical sense to perform partial trace operations over them. If we were doing partial traces, all Slater determinants would be maximally entangled due to the antisymmetry of the wave function. Fortunately, there are several ways to characterize and measure correlations in an indistinguishable system such as the fermion partial trace between modes \cite{fermionic_partial_trace,friis_relativity} or the von Neumann entropy of the one-body density matrix \cite{yoshiko,gigena_manybody_entanglement,gigena_overall_entropy}. Following the former, we will define the subsystems as the orbitals (or modes) instead of particles. In this way, we can treat the system as a tensor product of orbitals, if we take into account some subtleties which arise from the fermion anticommutation rules \cite{fermionic_partial_trace,friis_relativity}.

With this scheme in mind, correlations between modes can be purely classical, purely quantum, or a mixture of both \cite{ding,ding_2,caroline_robin}. The total correlation, which is the sum of the quantum and classical ones, is measured through the quantum version of the mutual information. If we divide our system into two parts ($A$ and $B$), the mutual information is defined as
\begin{equation}
    I(A,B) = S(\rho^{(A)})+S(\rho^{(B)})-S(\rho^{(A,B)})
\end{equation}
where $S(\rho^{(A)}) = -\Tr(\rho^{(A)}\ln\rho^{(A)})$ is the von Neumann entropy, $\rho^{(A)} = \Tr_B(\ket{\psi}\bra{\psi})$ is the reduced density matrix of the subsystem $A$ obtained through the fermion partial trace operation \cite{fermionic_partial_trace,friis_relativity} (the same applies for the subsystem $B$) and $\rho^{(A,B)}$ is the total density matrix. There are an equivalent alternative expression for the classical mutual information ($I(A,B) = S(A)+S(B)-S(A,B)$), given by
\begin{equation}
\label{eq:mutual_info_classical}
    I_{alt}(A,B) = S(A)-S(A|B) 
\end{equation}
where $S(A)$ is the entropy of the subsystem $A$ and $S(A|B)$ is the entropy of the subsystem $A$ conditioned to the information obtained by the subsystem $B$. In Quantum Physics, the information of a subsystem is obtained through projective measurements, so the quantum version of Eq. (\ref{eq:mutual_info_classical}) is \cite{quantum_discord,henderson}
\begin{equation}
\label{eq:J}
J(A,B) = \max_{\{\Pi_{k}^{(B)}\}} S(\rho^{(A)})-S(\rho^{(A,B)}|\{\Pi_{k}^{(B)}\}).
\end{equation}
While $I(A,B)$ is a measure of all kind of correlations, $J(A,B)$ quantifies only the classical part. The measurement-based conditional entropy in Eq. (\ref{eq:J}) is defined as
\begin{equation}
\label{eq:MBCE}
S(\rho^{(A,B)}|\{\Pi_{k}^{(B)}\}) = \sum_{k} p_k S(\rho_k^{(A,B)})
\end{equation}
where $\rho_k^{(A,B)} = \frac{1}{p_k}\Pi_{k}^{(B)}\rho^{(A,B)}\Pi_{k}^{(B)}$ is the measured-projected total state and $p_k = \tr (\Pi_{k}^{(B)}\rho^{(A,B)}\Pi_{k}^{(B)})$ is the associated probability. The measurement and the associated projector $\Pi_{k}^{(B)}$ are defined only in the sector $B$ of the bi-partition. With these definitions, one introduces the quantum discord \cite{quantum_discord,henderson,quantum_discord_review} as the difference between the total correlations and the classical ones
\begin{equation*}
\delta(A,B) = I(A,B) - J(A,B)
\end{equation*}
For pure states, the quantum discord reduces to entanglement \cite{luo}. However, this is no longer true for mixed states. This quantity is very useful in order to study quantum phase transitions in many-body systems  \cite{discord_phase_transitions,discord_phase_transitions_2,discord_phase_transitions_3}. Unfortunately, Eq. (\ref{eq:J}) requires a variational procedure involving all the possible $B$-subsystem projectors, so that computing quantum discord is in general intractable \cite{QD_is_NP}. Nevertheless, if we are dealing with fermion systems, no optimization process is required for the two orbital case because of the parity super-selection rule (PSSR) \cite{two_orbital_quantum_discord}.

In order to compute the quantum discord between more than two orbitals, the optimization process can be performed using a method similar to the one used by Shunlong Luo in \cite{luo} but taking into account some subtleties arising from the fermion nature of our system. If the subsystem $B$ is formed by $L_B$ orbitals, then we can parametrize the $B$-projectors $\Pi_k^{(B)}$ as
\begin{equation}
\label{eq:unitary_transformation}
    \Pi_k^{(B)} \rightarrow R^\dagger \Pi_k^{(B)} R
\end{equation}
for $k = 0,...,2^{L_B}-1$ with $R$ an unitary operator. However, not all the projectors parametrized in this way are allowed. Since we are dealing with fermion systems, the PSSR must be fulfilled \cite{ring_schuck,PSSR,friis_relativity}, so that the projected state has no coherence between different parity sectors. A way to achieve this is by using  Thouless rotations, i.e, $R = e^{iH}$ with 
\begin{equation}
\label{eq:Thouless_exponential}
    H = \sum_{ij\in \mathcal{H}_B}h_{ij}c^\dagger_ic_j + \frac{1}{2}\Delta_{ij}(c^\dagger_ic^\dagger_j+c_jc_i)
\end{equation}
an hermitian ($H = H^\dagger$) one body operator \cite{ring_schuck}. When the fermion operator $c^\dagger_i$ is transformed according to a rotation $R = e^{iH}$ with $H$ given by Eq. (\ref{eq:Thouless_exponential}), we obtain
\begin{equation*}
    Rc^\dagger_i R^\dagger = \sum_j U_{ji} c^\dagger_j + V_{ji}c_j = \beta^\dagger_i
\end{equation*}
where the parameters $U_{ij}$ and $V_{ij}$ (which define the new orbital basis) are related with the ones in Eq. (\ref{eq:Thouless_exponential}) by
\begin{equation*}
    \begin{pmatrix}
    U & V^* \\
    V & U^*
    \end{pmatrix}
    =
    \exp [i
    \begin{pmatrix}
    h & \Delta \\
    -\Delta^* & -h^*
    \end{pmatrix}
    ]
\end{equation*}
and $\beta^\dagger_i$ is a fermion creation operator. In this way, the parametrized projectors are projectors onto a state with well defined occupation in the orbital basis given by $\beta^\dagger_i$. The variational parameters\footnote{If we perform a Thouless rotation between $M$ orbitals, the number of variational parameters is $M(2M-1)$.} in Eq. (\ref{eq:J}) are $h_{ij}$ and $\Delta_{ij}$ with $h_{ij}=h_{ji}^*$ and $\Delta_{ij} = -\Delta_{ji}$. In order to find analytically the orbital basis that minimizes Eq. (\ref{eq:MBCE}), we can solve the equation $\delta S(\rho^{(A,B)}|\{\Pi_{k}^{(B)}\}) = 0$ where the variation of the measurement-based conditional entropy is given by the infinitesimal rotation $R \approx \mathbb{I} + iH$ (where $\mathbb{I}$ is the identity matrix). Then we obtain
\begin{align*}
    \sum_k & \ln (\langle \Pi_k\rangle)\langle \comm{\Pi_k}{H}\rangle - \langle\comm{\ln(\Pi_k\rho\Pi_k)}{H}\rangle = 0
\end{align*}
$\forall h_{ij}, \Delta_{ij}$, see Eq. (\ref{eq:Thouless_exponential}). The null eigenvalues of $\Pi_k\rho\Pi_k$ don't contribute to the sum. This equation is in general hard to solve analytically. However, the minimization process parametrized by Thouless rotations can be implemented in a computer in order to solve it numerically. This is applied in the next section, under the context of the three level Lipkin model.

Since we will compute quantum discord between two orbitals or two orbital pairs (as we will explain in Sec. \ref{sec:3_level_lipkin}), a Thouless rotation is enough in order to reach all the possible $B$-projectors $\Pi_k^{(B)}$. If we count the number of variational parameters in the case of $L_B=2$ for a general unitary respecting PSSR and a Thouless rotation, we obtain $6$ degrees of freedom for both\footnote{A general $N\times N$ unitary matrix has $N^2$ degrees of freedom. If we take into account the PSSR, we obtain $\frac{N^2}{2}$. However, since we are dealing with the transformation given in Eq. (\ref{eq:unitary_transformation}), the degrees of freedom corresponding to an overall phase and the decomposition $H = H' + a\mathbb{I}$ leave the transformation invariant, so we obtain $\frac{N^2}{2}-2$. For $N=2^{L_B}=4$, the number of variational parameters is $6$, which coincide with a Thouless rotation of $M=L_B=2$ orbitals.}. However, for a general case in which the subsystem $B$ is formed by $L_B>2$ orbitals, one may need more terms in Eq. (\ref{eq:Thouless_exponential}) such as two body fermionic operators and so on. This is an interesting topic for future work, but it is out of the scope of this article.

\section{\label{sec:3_level_lipkin}Three level Lipkin model}

\begin{figure}[h]
\includegraphics[width=0.5\textwidth]{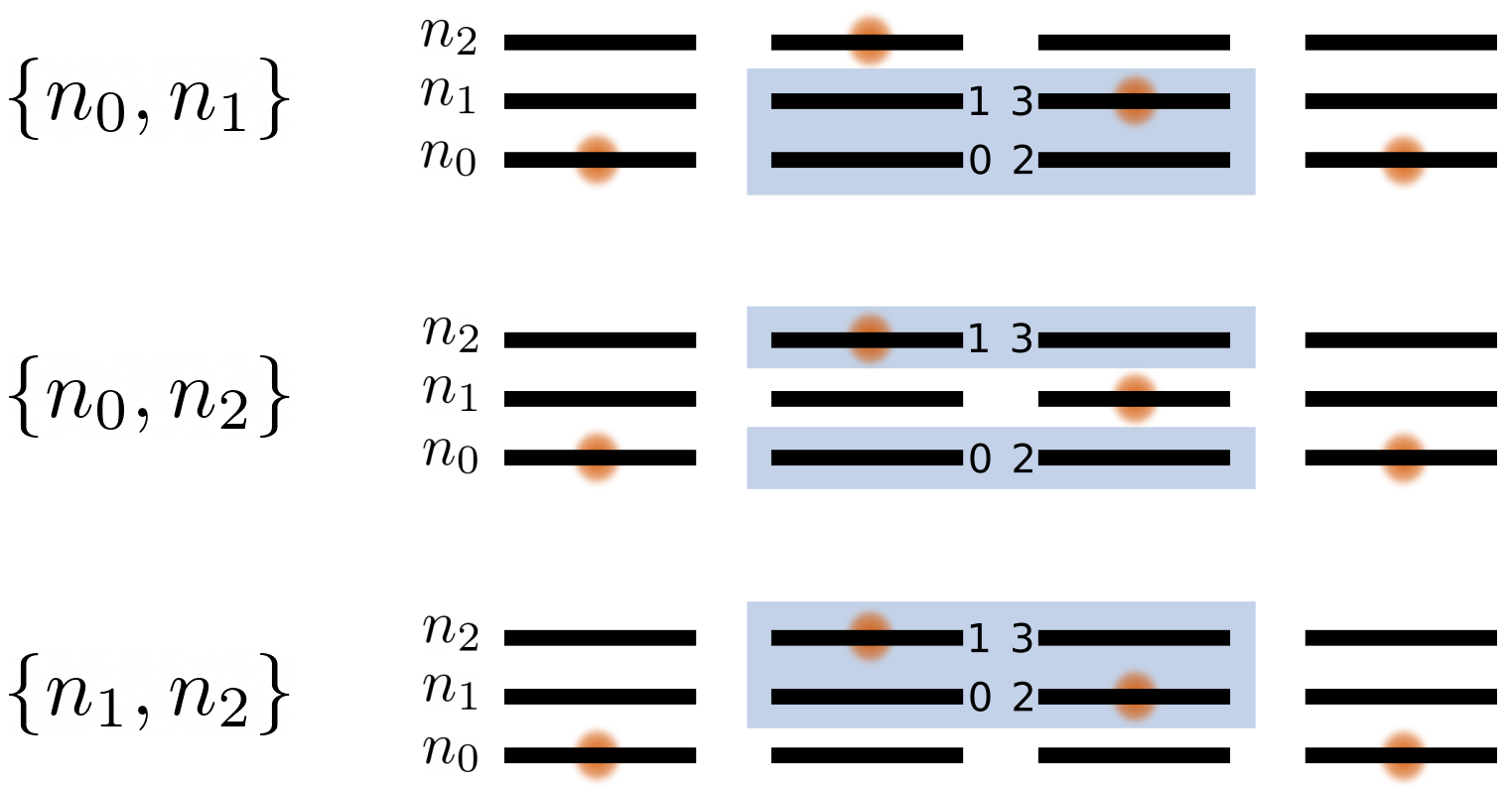}
\caption{Orbital numeration for the 4-orbital reduced density matrix in the three different subsystems (highlighted in blue) studied in this work. The numbers $n_0$, $n_1$ and $n_2$ refers to the population of the zeroth, first and second energy level respectively, while $\{n_0,n_1\}$, $\{n_0,n_2\}$ and $\{n_1,n_2\}$ refers to the orbital subsystems in blue. The orbitals within a subsystem are labeled according to the numbers in the blue region.}
\label{fig:3Lipkin_model_scheme}
\end{figure}

The three level Lipkin model is a generalization of the two level Lipkin model \cite{lipkin}. It contains three energy levels, each of them having a $N$-fold degeneracy \cite{HOLZWARTH,three_lipkin_shou,bertsch} and the dynamic evolution is governed by the Hamiltonian

\begin{equation}
\label{eq:3_level_Lipkin_Hamiltonian}
    H = \epsilon(K_{22}-K_{00})-\frac{V}{2}(K^2_{10}+K^2_{20}+K^2_{21}+h.c)
\end{equation}
with
\begin{equation*}
    K_{\sigma\sigma '} = \sum_{p=1}^N c^\dagger_{\sigma p}c_{\sigma ' p}
\end{equation*}
and $\sigma \in \{0,1,2\}$ labeling the three different energy levels. Fermion creation and annihilation operators, satisfying canonical anticommutation relations (CAR), are denoted by  $c^\dagger_{\sigma p}$ and $c_{\sigma p}$. The operators $K_{\sigma\sigma '}$ are the generators of the algebra of $SU(3)$ \cite{HOLZWARTH}, and one can use this property to compute numerically the exact ground state of the model 
\begin{equation}
\label{eq:exact_state}
    \ket{\psi} = \sum_{pq}C_{pq}\ket{pq}
\end{equation}
by using the basis $\ket{pq}$, where $p$ ($q$) is the number of particles in the first (second) energy level. The states $\ket{pq}$ are proportional to an equally weighted superposition of all possible states in the occupational basis that are eigenstates of $K_{00}$, $K_{11}$ and $K_{22}$ with eigenvalues $N-p-q$, $p$ and $q$ respectively (see Appendix \ref{sec:4_orbital_reduced_density_matrix} for more details). Thus, it is possible to easily obtain the reduced density matrix for a $4$-orbital subsystem in order to compute the quantum discord between two pairs of orbitals. Using the methods explained in Sec. \ref{sec:theory} and \cite{two_orbital_quantum_discord} we will compute the two orbital and two orbital pair quantum discord for the exact, HF, PHF and GCM ground states in different subsystems (illustrated in Fig. \ref{fig:3Lipkin_model_scheme}) and different values of the interaction parameter of the Hamiltonian ($V$ in Eq. (\ref{eq:3_level_Lipkin_Hamiltonian})). More details about the derivation of the HF, PHF and GCM solutions can be found in the Appendix \ref{sec:4_orbital_reduced_density_matrix} and \ref{sec:GCM}.

The three different subsystems, highlighted in Fig. \ref{fig:3Lipkin_model_scheme}, are the 4-orbital reduced density matrices corresponding to two orbitals of the $i$-th energy level and two orbitals of the $j$-th energy level. The two orbitals of the upper energy levels have the same degeneracy quantum numbers as the two orbitals of the lower energy level. The three 4-orbital subsystems are denoted by $\{n_i, n_j\}$ with $i,j = 0,1,2$ and $i\neq j$. For each $\{n_i, n_j\}$ subsystem, we label the four corresponding orbitals following the notation in Fig. \ref{fig:3Lipkin_model_scheme}: from left to right (i.e, from $p=1$ to $p=N$), the orbitals within the lowest energy level with the numbers $0,2$ and the ones within the highest energy level with $1,3$. For example, if we study correlations between two orbitals within the $2$nd energy level and two orbitals within the $0$th energy level, we will study the partition with $A = \{1,3\}$ and $B = \{0,2\}$ of the $\{n_0,n_2\}$ subsystem.

It is well known that the three level Lipkin  model shows two quantum phase transitions  (QPT) \cite{HOLZWARTH} as a function of the interaction parameter $\chi=\frac{V(N-1)}{\epsilon}$. These QPT are observed when the HF ground state develops spontaneous symmetry breaking of the parity quantum number corresponding to the second and third orbitals at $\chi=1$ and $\chi=3$, respectively. Within this context, we define the level's parity as the parity of the number of fermions that occupy that energy level.

Some entanglement and correlation properties of the ground state have been studied for the two level version \cite{entanglement_lipkin_model,Bao_2021} and in the thermodynamical limit \cite{Bao_2021,vidal_1,vidal_2}.

\subsection{\label{sec:QD_01}Quantum discord for the \{$n_0$, $n_1$\} subsystem}

\begin{figure}[h]
\includegraphics[width=0.5\textwidth]{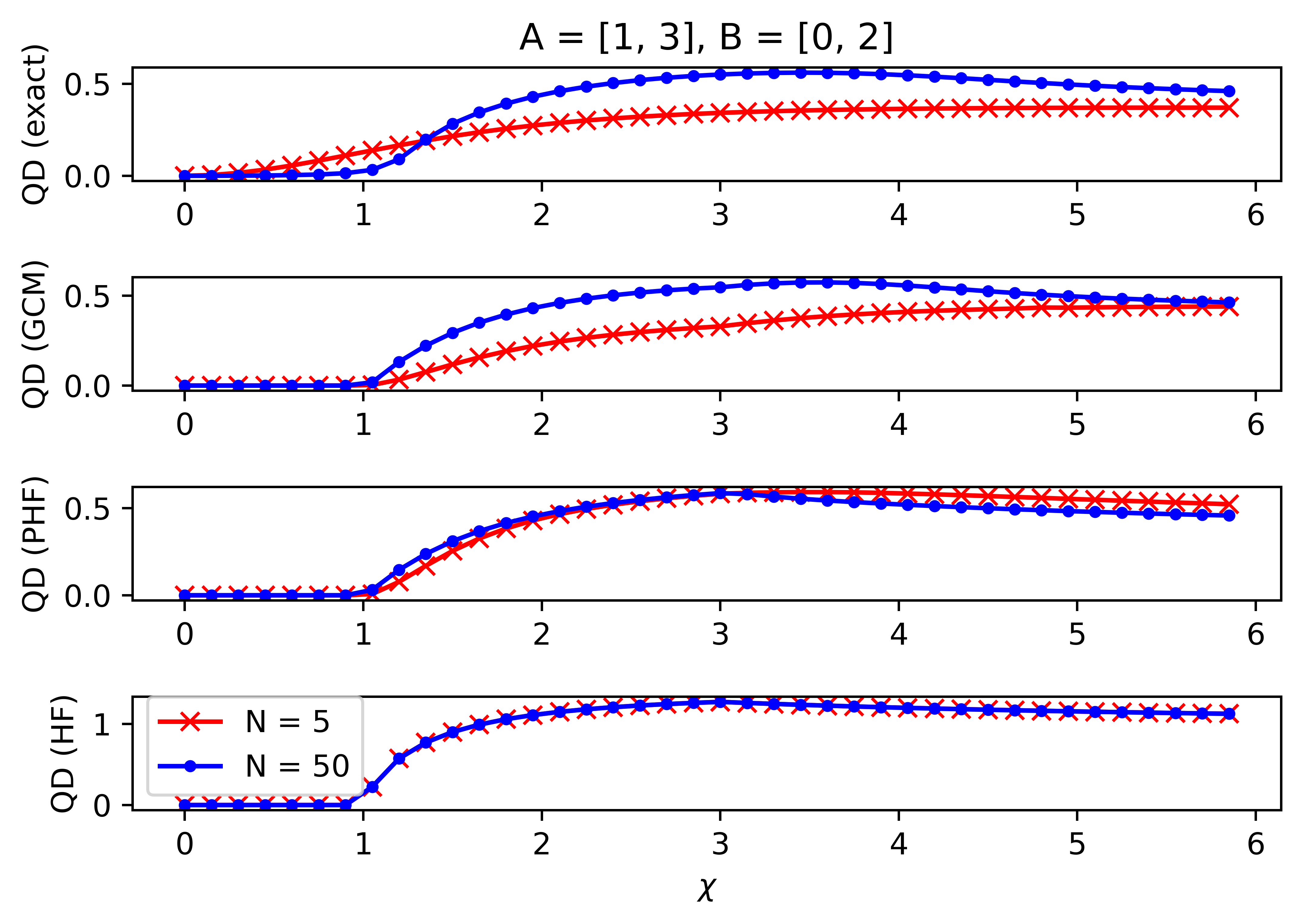}
\caption{Quantum discord for the partition given by $A=\{1,3\}$, $B=\{0,2\}$ for the exact (up), GCM (middle up), PHF (middle down) and HF (down) solutions.}
\label{fig:partition_01/A13_B02}
\end{figure}

One can apply the method explained in Sec. \ref{sec:theory} to compute the quantum discord for a given partition as a function of the interaction parameter $\chi$ (more details about the reduced density matrix can be found in Appendix \ref{sec:4_orbital_reduced_density_matrix}) and using the exact, HF, PHF and GCM solutions. In Fig. \ref{fig:partition_01/A13_B02} we represent the quantum discord between the partition given by $A = \{1,3\}$ and $B = \{0,2\}$ within the $\{n_0,n_1\}$ subsystem\footnote{The quantum discord is, in general, non symmetric. However, for this partition and all subsystems considered in this work, it has been found numerically that the exchange $A\leftrightarrow B$ leaves the quantum discord invariant.}.

In the HF case, we observe both QPTs through the change of behavior of the QD when $\chi = 1$ and $\chi = 3$. At $\chi = 1$, the QD suddenly increases as the HF state is not anymore in the non interacting ground state and the orbitals of the zeroth and first level start mixing. When $\chi = 3$, the mixing in the HF solution also include the second orbital, and the correlation is redistributed, so the QD stops increasing. The PHF solution is very similar to the HF one. Here, the QD slightly depends on the particle number and the projection to the good parity quantum numbers does not modify significantly the shape of the curve and therefore the discussion for the HF case still holds. On the other hand, the QD for the GCM solution is very accurate when $\chi\geq 1$ but identical to the HF and PHF solutions if $\chi\leq 1$. This makes sense, since the coordinate corresponding to the occupation of the first level is fixed to the corresponding HF value and the coordinate corresponding to the second level is treated by using the Hill-Wheeler equation (see Appendix \ref{sec:GCM} for more details).
Overall, all the approximate solutions within this particular partition have a good behavior in terms of QD, specially when the particle number is high.

\begin{figure}[h]
\includegraphics[width=0.5\textwidth]{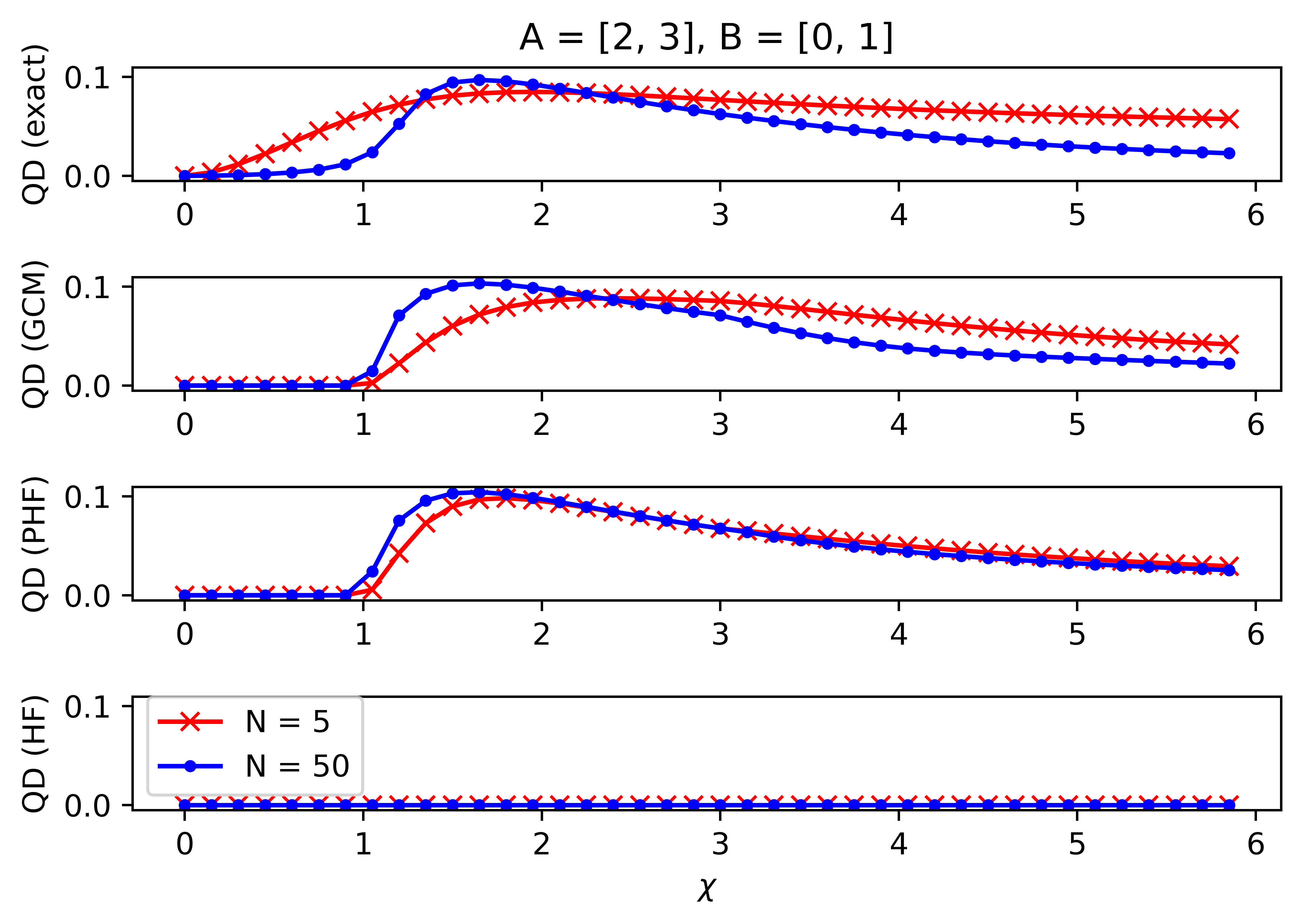}
\caption{Quantum discord for $A=\{2,3\}$ and $B=\{0,1\}$ in the exact (up), GCM (middle up), PHF (middle down) and HF (down) solutions.}
\label{fig:partition_01/A23_B01}
\end{figure}

However, while this approximations succeed in order to catch the behavior of quantum correlations between the $\{0, 2\}$ and $\{1, 3\}$ orbitals, this is not the case for other partitions. For example, if we set $A = \{2,3\}$ and $B = \{0,1\}$, we obtain Fig. \ref{fig:partition_01/A23_B01}.

\begin{figure}[h]
\includegraphics[width=0.5\textwidth]{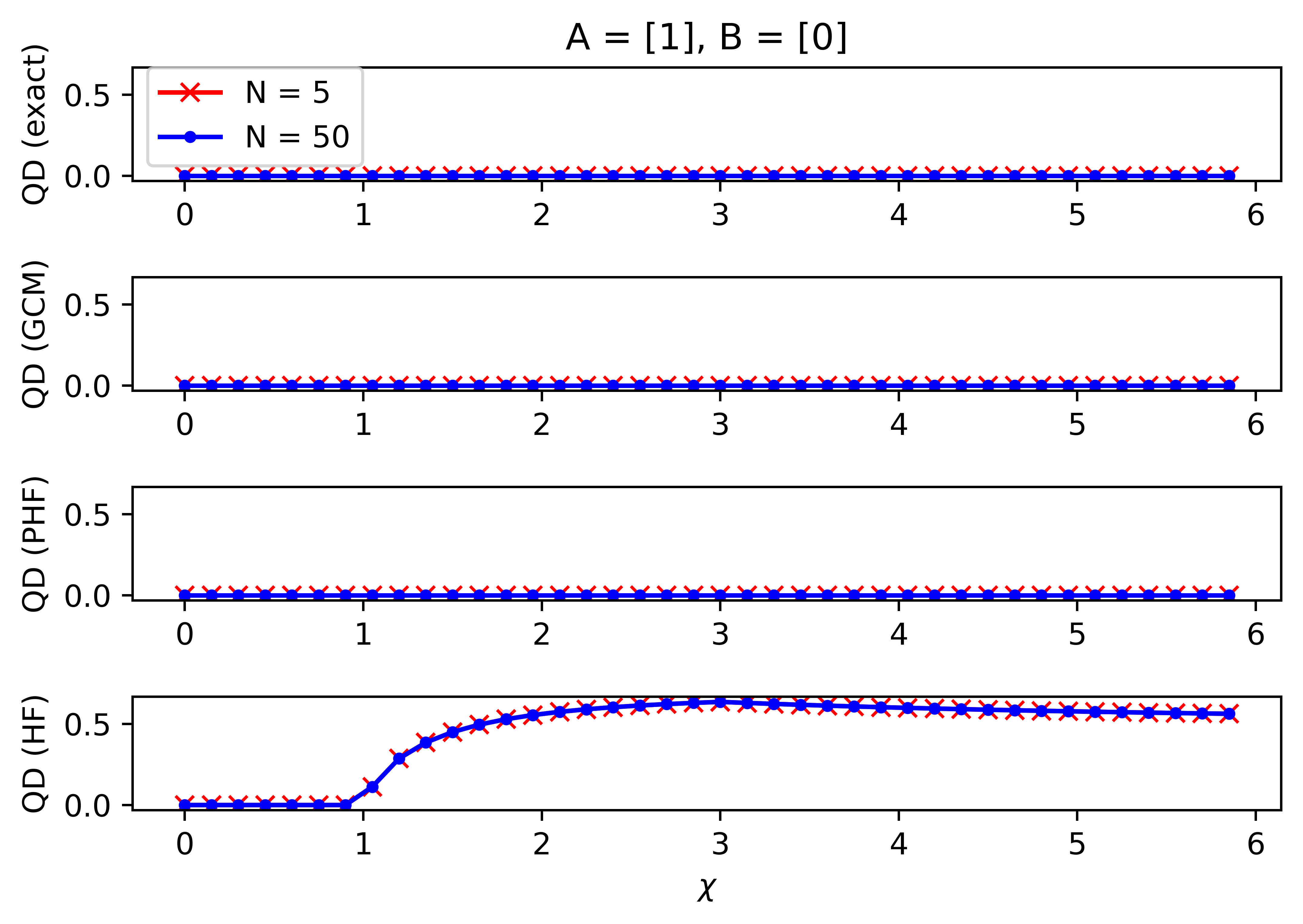}
\caption{Quantum discord for $A=\{1\}$ and $B=\{0\}$ in the exact (up), GCM (middle up), PHF (middle down) and HF (down) solutions.}
\label{fig:partition_01/A1_B0}
\end{figure}

We observe a clearly different behavior between the HF ground state and all the other ones. Since the HF orbitals do not mix states with different degeneration number, the QD is zero for all values of $\chi$. However the PHF and GCM solutions are successful capturing the quantum correlations from the exact state. As in Fig. \ref{fig:partition_01/A13_B02}, the PHF solution slightly depends on the particle number and the GCM solution is closer to the exact one specially if the particle number is large. However, since the coordinate for the first level is fixed to the HF solution, the GCM ground state is not able to reproduce accurately the QD behavior when $\chi\leq 1$. The HF solution fails to capture the quantum correlations of the exact state. The opposite behavior takes place when we analyze the two orbital QD for the $A=\{1\}$, $B=\{0\}$ partition (Fig. \ref{fig:partition_01/A1_B0}).

Due to conservation of parity symmetry, the exact ground state does not have any quantum correlations, as well as the GCM and PHF solutions. The two orbital quantum discord for fermion systems with definite number of particles depends only on the off-diagonal elements of the one body density \cite{two_orbital_quantum_discord} and in this case they are zero because of the parity symmetry. However, the HF solution breaks the parity symmetry, and fake quantum correlations appear since the off-diagonal elements of the corresponding one body density matrix are in general non zero. Exactly the same behavior is observed for the two orbital quantum discord within the $\{n_0,n_2\}$ and $\{n_1,n_2\}$ subsystems (Fig. \ref{fig:partition_02/A1_B0}), and partitions $A = \{0,2\}$, $B = \{1\}$ or $A = \{1,3\}$, $B = \{0\}$.

\subsection{Quantum discord for the \{$n_0$, $n_2$\} subsystem}

\begin{figure}[h]
\includegraphics[width=0.5\textwidth]{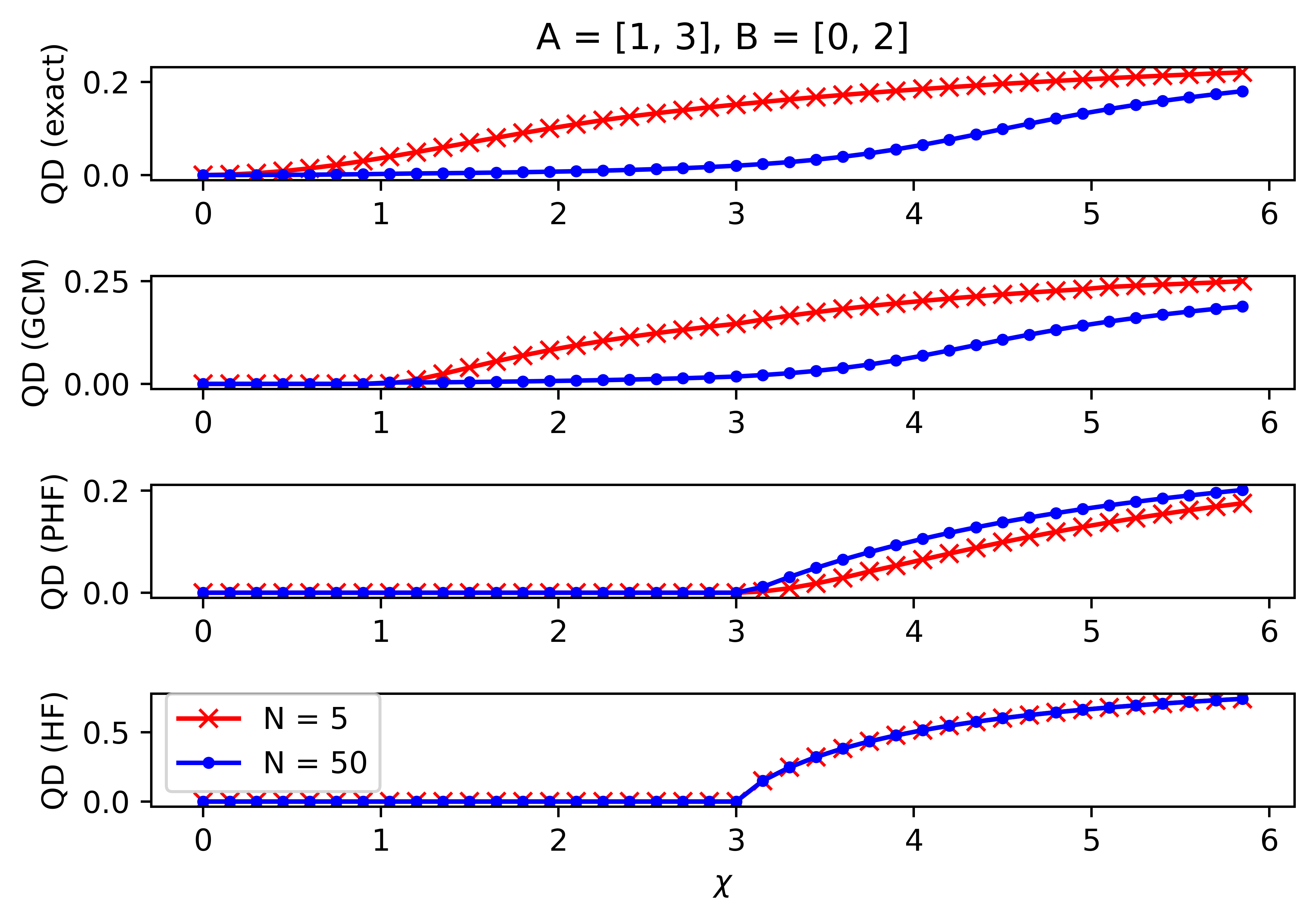}
\caption{Quantum discord for $A=\{1,3\}$ and $B=\{0,2\}$ in the exact (up), GCM (middle up), PHF (middle down) and HF (down) solutions.}
\label{fig:partition_02/A13_B02}
\end{figure}

The results obtained for this subsystem, which is schematically represented in Fig. \ref{fig:3Lipkin_model_scheme}, are in general similar to the results obtained for the subsystem $\{n_0,n_1\}$. In Fig. \ref{fig:partition_02/A13_B02} we represent the quantum discord for the partition $A = \{1,3\}$, $B = \{0,2\}$.

Since the HF orbitals do not have components from the second energy level orbitals until $\chi\geq 3$, the zeroth level orbitals remain uncorrelated with the second ones until the second QPT for the HF and PHF solutions takes place. This is not the case for the GCM ground state because only the first coordinate (which corresponds to the first energy level) is fixed to the value of the HF solution. This is reflected in the behavior of the QD when $1\leq\chi\leq 3$: while the exact and GCM solutions have a non zero QD, the HF and PHF ground states are not correlated. The reason why the GCM approximation is practically identical to the exact solution in this case, is that the QD for the exact state is very small until $\chi \approx 1$. This is in total agreement with the GCM solution, which shows zero QD until $\chi = 1$. Again, the accuracy of the approximate methods is better if the particle number is large.

\begin{figure}[h]
\includegraphics[width=0.5\textwidth]{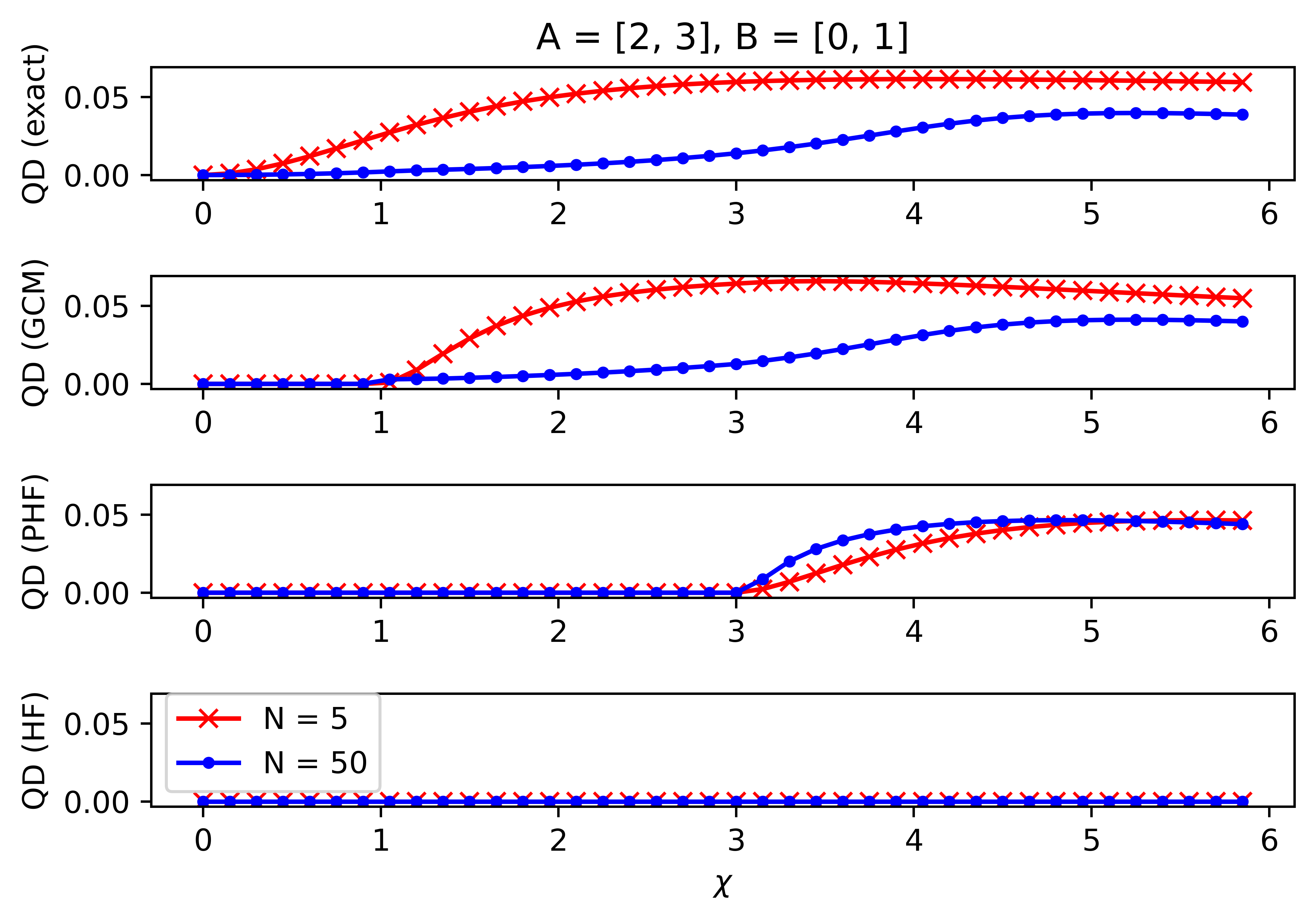}
\caption{Quantum discord for $A=\{2,3\}$ and $B=\{0,1\}$ in the exact (up), GCM (middle up), PHF (middle down) and HF (down) solutions.}
\label{fig:partition_02/A23_B01}
\end{figure}

Let us now consider the $A = \{2,3\}$ and $B = \{0,1\}$ partition in the $\{n_0,n_2\}$ subsystem. In this case we obtain Fig. \ref{fig:partition_02/A23_B01} where we observe that, again the HF state does not catch any quantum correlations because of the same reasons as in Fig. \ref{fig:partition_01/A23_B01}: the HF orbitals do not mix states with different degeneration number. The PHF state, which is no longer a Slater determinant, reproduce approximately the behavior of the exact QD for $\chi\geq 3$. However, since it is constructed through the projection of the HF state, the QD for $\chi\leq 3$ is zero (the second level orbitals are not populated until $\chi = 3$). As in the previous situations, the QD depends on the particle number, weakly for the PHF solution and dramatically for the GCM state. For the later, the same discussion as in the $A=\{1,3\}$ and $B=\{0,2\}$ partition applies: the region $\chi\leq 1$ is not well approximated when the particle number is small because of the HF coordinate of the first energy level.

\begin{figure}[h]
\includegraphics[width=0.5\textwidth]{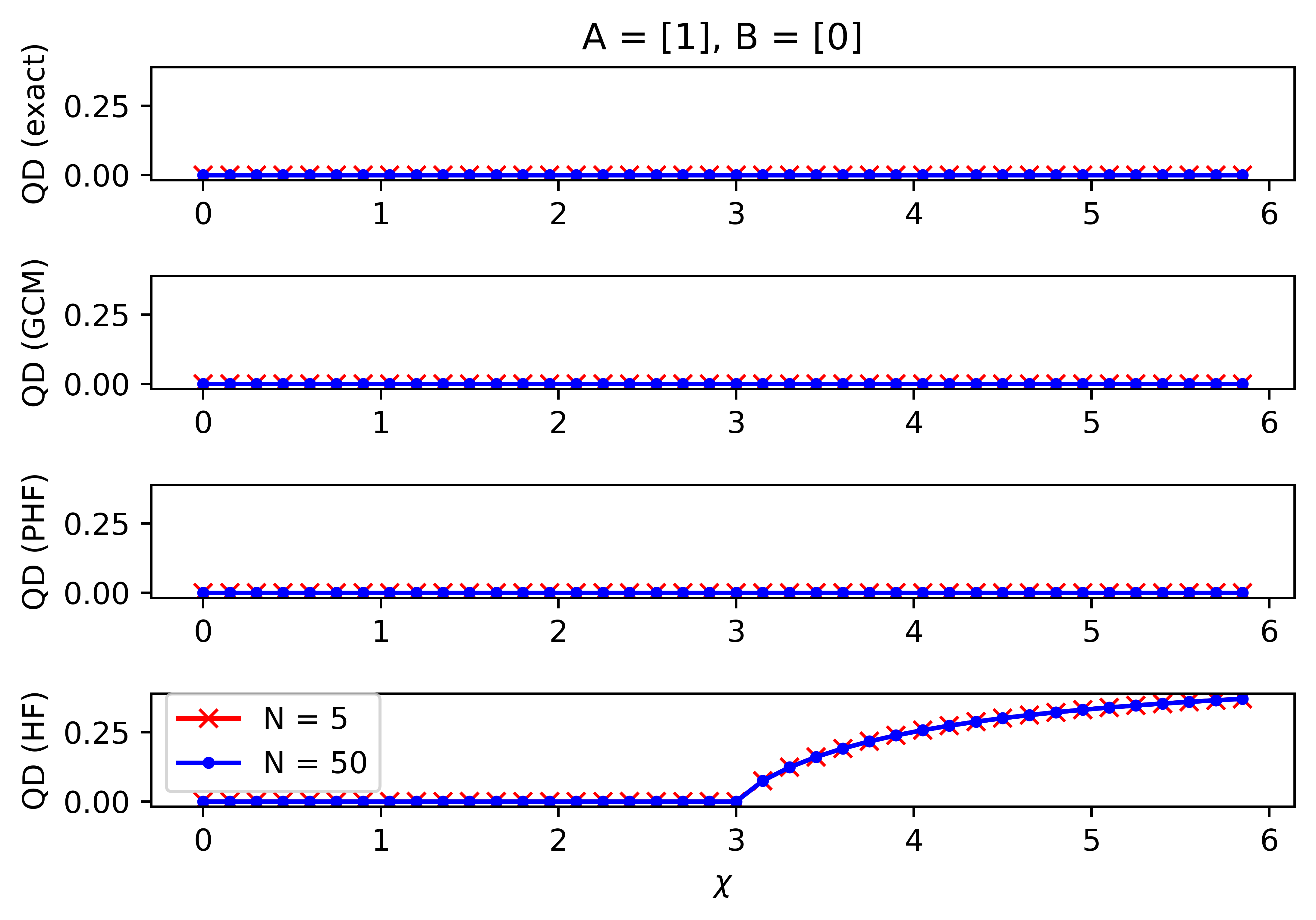}
\caption{Quantum discord for $A=\{1\}$ and $B=\{0\}$ in the exact (up), GCM (middle up), PHF (middle down) and HF (down) solutions.}
\label{fig:partition_02/A1_B0}
\end{figure}

Finally, the quantum discord between two orbitals (Fig. \ref{fig:partition_02/A1_B0}) acquires nonzero values only for the Hartree-Fock state and the same discussion as in Sec. \ref{sec:QD_01} applies here. We obtain the same results for $A = \{1,3\}$, $B = \{0\}$ and $A = \{2,0\}$, $B = \{1\}$.

The results obtained for the \{$n_1,n_2$\} subsystem are very similar and the discussion is essentially the same, so they are presented in Appendix \ref{sec:n_1__n_2__partition} for completeness.

The behavior of the variational parameters of the Thouless rotation in Eq. (\ref{eq:Thouless_exponential}) depends on the partition and subsystem. Since the rotations performed in this work are always between two orbitals, the matrices given by the elements $h_{ij}$ and $\Delta_{ij}$ will only have one off-diagonal element. In this way, it is interesting to remark the behavior of the quantities $|h_{ij}|^2$ and $|\Delta_{ij}|^2$ for $i\neq j$. For all partitions and subsystems, the value of the quantum discord does not depend on the quantity $|\Delta_{ij}|^2$. For $|h_{ij}|^2$ we observe the same behavior for the partition $A = \{1,3\}$ $B=\{0,2\}$, but not for $A = \{2,3\}$ $B=\{0,1\}$. In the latter, we see that $|h_{ij}|^2$ acquires different approximately constant values (for the non-exact solutions) whose numerical values depend on whether the interaction parameter $\chi$ has reached the critical value where the quantum phase transition happens ($\chi_c=1,3$) or no, and dividing the parameter space into three different regions: $\chi < 1$, $1<\chi<3$ and $\chi > 3$.


\section{\label{sec:conclusions}Discussion and conclusions}


We have carried out an analysis of quantum correlations for the three level Lipkin model, which is a solvable but non trivial model that manifests two quantum phase transitions due to two spontaneous symmetry breaking processes, corresponding to the parity symmetry for the first and second energy levels. We have computed the quantum discord, which is a well known measure of a system's quantum correlations, including entanglement and beyond. Since we are dealing with fermion systems composed by orbitals, we have carried out the minimization process of the measurement-based conditional entropy through Thouless rotations. In this way we make sure that the projectors don't break the parity superselection rule and that they have a clear physical sense. More precisely, they correspond to projections to the occupation of some (quasi)particle orbitals. We have analyzed the quantum discord as a function of the interaction parameter of the model $\chi$, within different subsystems (illustrated in Fig. \ref{fig:3Lipkin_model_scheme}).


The partition corresponding to $A = \{1,3\}$, $B = \{0,2\}$ is, in some way, the natural partition given the interaction of the Hamiltonian (Eq. \ref{eq:3_level_Lipkin_Hamiltonian}). The operators $K_{ij}^2$ with $i\neq j$, when applied to a Slater determinant\footnote{That Slater determinant must be invariant with respect to permutations of orbitals with different degeneration number within the same energy level.}, annihilate a pair of particles within the $i$-th level and create another pair within the $j$-th level in a symmetric superposition such as the resulting state is invariant with respect to permutations of the degeneration orbitals (within the same energy level). Thus, one would expect quantum correlations between $A = \{1,3\}$, $B = \{0,2\}$ to be a good measure of the interaction. In practice we observe that, for this `interaction partition', all approximated solutions reflect (with different accuracy) the behavior of the exact result. However, for other partitions such as $A = \{2,3\}$, $B = \{0,1\}$ or $A = \{1\}$, $B = \{0\}$, the HF solution does not catch the QD corresponding to the exact state. This is related to the QPTs and the symmetry breaking mechanism.

The QPTs play a significant role in all results presented. When the system undergoes a QPT, there is a change of behavior in the QD depending on the partition. If the QPT is related to  symmetry breaking of the $i$-th energy level, the QD of a partition within the corresponding subsystem suddenly grows. This behavior is sharper when the particle number is large. However, for the HF solution, this behavior is only present when the partition is the `interaction partition'. For the $A = \{2,3\}$, $B = \{0,1\}$ partition, the HF does not catch quantum correlations because of the way the HF orbitals are defined. Since they only mix orbitals with the same degeneration number, it is expected that a partition whose QD measures the quantum correlations between pairs of orbitals with different degeneration number has zero QD for all $\chi$. On the other hand, the absence of parity symmetry of the HF solution causes a non zero QD of the two orbital partition $A = \{1\}$, $B = \{0\}$, since transitions between single orbitals are allowed. Those `fake quantum correlations'\footnote{We call it `fake' because they are not present in the exact solution.} are possible because the HF orbitals are defined as linear combinations of orbitals from different energy levels with same degeneration, which is the same reason by which the QD in the $A = \{2,3\}$, $B = \{0,1\}$ vanishes. With this in mind, it is reasonable to think about the symmetry breaking process as a `redistribution' of the quantum correlations within a system, from higher level to lower level ones\footnote{Here, we refer to `high level correlations' to the ones which involve bigger partitions, and `low level correlations' to the ones which involve smaller partitions.}.

The symmetry restoration process recovers (or redistributes) the correct correlations between the different partitions, and it succeeds specially when the particle number is large and far from the QPT (i.e, when the interaction is high enough). This behavior is clearly observed in the GCM solution, where one coordinate (the one corresponding to the first energy level) is kept fixed to the HF value and the second one is used as coordinate in the GCM method.

Our results could help to investigate the correlation structure of more complex models, such as the Agassi model \cite{agassi1968}, which is a two level Lipkin model with a pairing interaction and a superconducting phase. Other type of mathematically motivated approximations, such as the `nearest' Slater determinant, defined in \cite{grassmannian}, could also be of interest.

\acknowledgments{}
The authors want to thank the Madrid regional government, Comunidad 
Aut\'onoma de Madrid, for the project Quantum Information Technologies: 
QUITEMAD-CM P2018/TCS-4342.
The  work of LMR was supported by Spanish Ministry 
of Economy and Competitiveness (MINECO) Grants No. 
PGC2018-094583-B-I00.

\appendix


\section{\label{sec:4_orbital_reduced_density_matrix}Four orbital reduced density matrix for exact, HF and PHF ground state}


In this appendix we derive the analytical expression for the four orbital reduced exact and HF ground state of the three level Lipkin model. The exact ground state of the three level Lipkin model can be expanded as \cite{HOLZWARTH} 
\begin{equation*}
    \ket{\psi} = \sum_{0\leq p+q\leq N} C_{pq}\ket{pq}
\end{equation*}
with 
\begin{equation*}
    \ket{pq} = \sqrt{\frac{(N-p-q)!p!q!}{N!}}\ket{n_1 = p, n_2 = q}
\end{equation*}
and
\begin{equation*}
    \ket{n_1n_2} = \sum_{\substack{n_1<...<n_p \neq \\ m_1<...<m_q}}^N \ket{1_{n_1}...1_{n_p}2_{m_1}...2_{m_q}}
\end{equation*}
The state $\ket{1_{n_1}...1_{n_p}2_{m_1}...2_{m_q}}$ introduced above corresponds to a Slater determinant with $p$ ($q$) particles occupying the first (second) energy level of the Lipkin model following the same-level distribution given by the indices $n_i$ ($m_j$). The coefficients $C_{pq}$ are determined by the diagonalization of the Hamiltonian matrix in the $\{\ket{pq}\}$ basis:
\begin{widetext}
\begin{align*}
    \bra{p'q'}H\ket{pq} & = \epsilon(2q+p-N)\delta_{p',p}\delta_{q',q} -\frac{V}{2}\bigg(\sqrt{(N-p-q+1)(N-p-q+2)p(p-1)}\delta_{p',p-2}\delta_{q',q} \\
    & + \sqrt{(N-p-q+1)(N-p-q+2)q(q-1)}\delta_{p',p}\delta_{q',q-2} + \sqrt{(p+1)(p+2)q(q-1)}\delta_{p',p+2}\delta_{q',q-2} \\
    & + \sqrt{(N-p-q-1)(N-p-q)(p+1)(p+2)}\delta_{p',p+2}\delta_{q',q} \\ 
    & + \sqrt{(N-p-q-1)(N-p-q)(q+1)(q+2)}\delta_{p',p}\delta_{q',q+2} + \sqrt{(q+1)(q+2)p(p-1)}\delta_{p',p-2}\delta_{q',q+2}
    \bigg)
\end{align*}
\end{widetext}
In the same manner, the HF ground state for this model can be written as
\begin{equation*}
    \ket{\psi_{HF}} = \prod_{i=1}^N a^\dagger_{0,i}\ket{\phi} = \sum_{n_1+n_2 \leq N}C_{n_1n_2}^{(HF)}\ket{n_1n_2}
\end{equation*}
where the coefficients $C_{n_1n_2}^{(HF)}$ are given by the relationship between the HF ($a^\dagger_{\sigma,i}$) and basis  ($c^\dagger_{\sigma,i}$) orbitals  (see Eq. \ref{eq:change_of_basis} defined later) \cite{HOLZWARTH}
\begin{align*}
    &C_{n_1n_2}^{(HF)} = \\
    &\begin{cases}
    1, & \text{if } n_1=n_2=0\text{ and } \chi \leq 1\\
    (\frac{1}{\sqrt{2}})^N(1+\frac{1}{\chi})^{\frac{N-n_1}{2}}(1-\frac{1}{\chi})^{\frac{n_1}{2}}, & \text{if } n_2=0\text{ and } 1\leq\chi\leq 3\\
    (\frac{1}{\sqrt{3}})^{n_1}(\frac{\chi+3}{3\chi})^{\frac{N-n_1-n_2}{2}}(\frac{\chi-3}{3\chi})^{\frac{n_2}{2}}, & \text{if } \chi\geq 3\\
    0, & \text{otherwise }\\
    \end{cases}
\end{align*}
In order to obtain the PHF ground state, we must project the HF state to obtain a wave function with the proper quantum numbers. For the three level Lipkin model, the wave function with the proper quantum numbers is the one with positive (even) parity of both the first and second energy levels, and definite total particle number. This symmetry restored wave function  can be obtained from the HF one through the projection operators
\begin{equation*}
    P_{+}^{(\sigma)} = \frac{1}{2}(\mathbb{I}+e^{i\pi K_{\sigma\sigma}})
\end{equation*}
with $\sigma = 1,2$. Since the resulting state will simply be the the original HF ground state with $C_{n_1n_2}^{(HF)} = 0$ if $n_1$ or $n_2$ are odd (and the nonzero coefficients properly normalized), we won't give the explicit expression of the reduced state.
For the exact and HF cases, the reduced four orbital density matrix for a given partition $\{n_i,n_j\}$ (see Fig. \ref{fig:3Lipkin_model_scheme}) is
\begin{align*}
    \rho^{(4)}_{\{n_i,n_j\}} =& \Tr_{\neg\{n_i,n_j\}}(\ket{\psi}\bra{\psi}) \\
    \rho^{(4,HF)}_{\{n_i,n_j\}} =& \Tr_{\neg\{n_i,n_j\}}(\ket{\psi_{HF}}\bra{\psi_{HF}}) 
\end{align*}
The operation $\Tr_{\neg X}(O)$ represents the fermion partial trace \cite{fermionic_partial_trace} of the operator $O$ and $\neg X$ is the subsystem that is traced out (since the symbol $\neg$ represents negation, $\Tr_{\neg X}(O)$ is the fermion partial trace operation over the orthogonal complement of $X$). With this results, it is straightforward to get analytical expressions for the four orbital reduced density matrix $\rho^{(4)}_{ij}$: 

\begin{widetext}
\begin{align}\label{eq:01_RDM}
    \rho^{(4)}_{\{n_0,n_1\}} &=  \frac{1}{N(N-1)}\sum_{p+q\leq N} \abs{C_{p,q}}^2\bigg\{ q(q-1)\ket{-,-}\bra{-,-}+q(N-p-q)\Big(\ket{-,0}\bra{-,0}+ \ket{0,-}\bra{0,-}\Big) \\
    & + pq\Big(\ket{-,1}\bra{-,1}+\ket{1,-}\bra{1,-}\Big) + (N-p-q)(N-p-q-1)\ket{0,0}\bra{0,0} \\
    & + p(N-p-q)\Big(\ket{0,1}\bra{0,1}+\ket{1,0}\bra{1,0}\Big) + p(p-1)\ket{1,1}\bra{1,1}
    \bigg\} \\
    & + \bigg\{C_{p,q}C_{p+2,q}^* \sqrt{(N-p-q)(N-p-q-1)(p+1)(p+2)}\ket{0,0}\bra{1,1} \\
    & + \abs{C_{p,q}}^2p(N-p-q)\ket{0,1}\bra{1,0} + h.c\bigg\},
\end{align}
\begin{align}\label{eq:02_RDM}
    \rho^{(4)}_{\{n_0,n_2\}} &=  \frac{1}{N(N-1)}\sum_{p+q\leq N} \abs{C_{p,q}}^2\bigg\{ p(p-1)\ket{-,-}\bra{-,-}+p(N-p-q)\Big(\ket{-,0}\bra{-,0}+ \ket{0,-}\bra{0,-}\Big) \\
    & + pq\Big(\ket{-,1}\bra{-,1}+\ket{1,-}\bra{1,-}\Big) + (N-p-q)(N-p-q-1)\ket{0,0}\bra{0,0} \\
    & + q(N-p-q)\Big(\ket{0,1}\bra{0,1}+\ket{1,0}\bra{1,0}\Big) + q(q-1)\ket{1,1}\bra{1,1}
    \bigg\} \\
    & + \bigg\{C_{p,q}C_{p,q+2}^* \sqrt{(N-p-q)(N-p-q-1)(q+1)(q+2)}\ket{0,0}\bra{1,1} \\
    & + \abs{C_{p,q}}^2q(N-p-q)\ket{0,1}\bra{1,0} + h.c\bigg\},
\end{align}
and
\begin{align}\label{eq:12_RDM}
    \rho^{(4)}_{\{n_1,n_2\}} &=  \frac{1}{N(N-1)}\sum_{p+q\leq N} \abs{C_{p,q}}^2\bigg\{ (N-p-1)(N-p-q-1)\ket{-,-}\bra{-,-}+p(N-p-q)\Big(\ket{-,0}\bra{-,0}+ \ket{0,-}\bra{0,-}\Big) \\
    & + q(N-p-q)\Big(\ket{-,1}\bra{-,1}+\ket{1,-}\bra{1,-}\Big) + p(p-1)\ket{0,0}\bra{0,0} \\
    & + pq\Big(\ket{0,1}\bra{0,1}+\ket{1,0}\bra{1,0}\Big) + q(q-1)\ket{1,1}\bra{1,1}
    \bigg\} \\
    & + \bigg\{C_{p+2,q}C_{p,q+2}^* \sqrt{(p+1)(p+2)(q+1)(q+2)}\ket{0,0}\bra{1,1} + \abs{C_{p,q}}^2pq\ket{0,1}\bra{1,0} + h.c\bigg\}
\end{align}
\end{widetext}
where the reduced states $\{\ket{\sigma_1,\sigma_2}\}$ with $\sigma_i = -,0,1$ refer to a) no occupation ($\sigma_i = -$) b) only the first level occupied ($\sigma_i = 0$) and c) only the second level occupied ($\sigma_i = 1$). The notation is used for both the $\{0,1\}$ orbitals ($\sigma_1$) and the $\{2,3\}$ orbitals ($\sigma_2$) and the orbital numeration in Fig. \ref{fig:3Lipkin_model_scheme} is followed for each subsystem. Similarly, we can compute analytical expressions for the four orbital Hartree-Fock reduced ground state $\rho^{(4,HF)}_{ij}$:
\begin{widetext}
\begin{align*}
    \rho^{(4,HF)}_{\{n_0,n_1\}} &=  U_{02}^4\ket{-,-}\bra{-,-} + U_{01}^2U_{02}^2\bigg(\ket{1,-}\bra{1,-}+\ket{-,1}\bra{-,1}\bigg) + U_{01}^4\ket{1,1}\bra{1,1} \\
    + & U_{00}^2U_{01}^2\bigg(\ket{1,0}\bra{1,0}+\ket{0,1}\bra{0,1}+\ket{0,1}\bra{1,0}+\ket{1,0}\bra{0,1}+\ket{0,0}\bra{1,1}+\ket{1,1}\bra{0,0}\bigg) \\
    + & U_{00}^4\ket{0,0}\bra{0,0} + U_{00}^2U_{02}^2\bigg(\ket{0,-}\bra{0,-}+\ket{-,0}\bra{-,0}\bigg) \\
    + & U_{00}U_{01}U_{02}^2\bigg(\ket{1,-}\bra{0,-}+\ket{-,0}\bra{-,1}+h.c\bigg) + U_{00}U_{01}^3\bigg(\ket{1,1}\bra{0,1}+\ket{1,0}\bra{1,1}+h.c\bigg) \\
    + & U_{00}^3U_{01}\bigg(\ket{0,0}\bra{0,1}+\ket{0,0}\bra{1,0}+h.c\bigg),
\end{align*}
\begin{align*}
    \rho^{(4,HF)}_{\{n_0,n_2\}} &=  U_{01}^4\ket{-,-}\bra{-,-} + U_{01}^2U_{02}^2\bigg(\ket{1,-}\bra{1,-}+\ket{-,1}\bra{-,1}\bigg) + U_{02}^4\ket{1,1}\bra{1,1} \\
    + & U_{00}^2U_{02}^2\bigg(\ket{1,0}\bra{1,0}+\ket{0,1}\bra{0,1}+\ket{0,1}\bra{1,0}+\ket{1,0}\bra{0,1}+\ket{0,0}\bra{1,1}+\ket{1,1}\bra{0,0}\bigg) \\
    + & U_{00}^4\ket{0,0}\bra{0,0} + U_{00}^2U_{01}^2\bigg(\ket{0,-}\bra{0,-}+\ket{-,0}\bra{-,0}\bigg) \\
    + & U_{00}U_{02}U_{01}^2\bigg(\ket{1,-}\bra{0,-}+\ket{-,0}\bra{-,1}+h.c\bigg) + U_{00}U_{02}^3\bigg(\ket{1,1}\bra{0,1}+\ket{1,0}\bra{1,1}+h.c\bigg) \\
    + & U_{00}^3U_{02}\bigg(\ket{0,0}\bra{0,1}+\ket{0,0}\bra{1,0}+h.c\bigg),
\end{align*}
and
\begin{align*}
    \rho^{(4,HF)}_{\{n_1,n_2\}} &=  U_{00}^4\ket{-,-}\bra{-,-} + U_{00}^2U_{02}^2\bigg(\ket{1,-}\bra{1,-}+\ket{-,1}\bra{-,1}\bigg) + U_{02}^4\ket{1,1}\bra{1,1} \\
    + & U_{02}^2U_{01}^2\bigg(\ket{1,0}\bra{1,0}+\ket{0,1}\bra{0,1}+\ket{0,1}\bra{1,0}+\ket{1,0}\bra{0,1}+\ket{0,0}\bra{1,1}+\ket{1,1}\bra{0,0}\bigg) \\
    + & U_{01}^4\ket{0,0}\bra{0,0} + U_{00}^2U_{01}^2\bigg(\ket{0,-}\bra{0,-}+\ket{-,0}\bra{-,0}\bigg) \\
    + & U_{02}U_{01}U_{00}^2\bigg(\ket{1,-}\bra{0,-}+\ket{-,0}\bra{-,1}+h.c\bigg) + U_{01}U_{02}^3\bigg(\ket{1,1}\bra{0,1}+\ket{1,0}\bra{1,1}+h.c\bigg) \\
    + & U_{01}^3U_{02}\bigg(\ket{0,0}\bra{0,1}+\ket{0,0}\bra{1,0}+h.c\bigg)
\end{align*}
\end{widetext}
where the coefficients $U_{ij}$ are the matrix elements of the change of basis matrix between the basis orbitals and the Hartree-Fock ones, i.e,
\begin{equation}
\label{eq:change_of_basis}
    a^\dagger_{\alpha,i} = \sum_{\beta}U_{\alpha\beta}c^\dagger_{\beta, i}
\end{equation}
with
\begin{align*}
    & U_{00}^2 = 
    \begin{cases}
    1 & \text{if } \chi\leq 1 \\
    \frac{1}{2}(1+\frac{1}{\chi}) & \text{if } 1\leq\chi\leq 3 \\
    \frac{\chi+3}{3\chi} & \text{if } \chi\geq 3 \\
    \end{cases}
    \\
    & U_{01}^2 = 
    \begin{cases}
    0 & \text{if } \chi\leq 1 \\
    \frac{1}{2}(1-\frac{1}{\chi}) & \text{if } 1\leq\chi\leq 3 \\
    \frac{1}{3} & \text{if } \chi\geq 3 \\
    \end{cases}
    \\
    & U_{02}^2 = 
    \begin{cases}
    0 & \text{if } \chi\leq 3 \\
    \frac{\chi-3}{3\chi} & \text{if } \chi\geq 3 \\
    \end{cases}
\end{align*}
With all this, we can compute easily the quantum discord with the method explained in Sec. \ref{sec:theory} for an arbitrary partition within the four orbital reduced state, both exact, mean-field or mean-field projected.


\section{\label{sec:GCM}GCM applied to the three level Lipkin model}


As the two level Lipkin model case \cite{robledo}, the exact solution for the three level Lipkin model can be obtained by means of the GCM method. More precisely, if we use as a generating functions the Slater determinants given by
\begin{equation}
\label{eq:SD_3Lipkin}
    \ket{\phi_1\phi_2} = \mathcal{N}\exp\big(\tan{\phi_1}(\cos{\phi_2}K_{10}+\sin{\phi_2}K_{20})\big)\ket{0}
\end{equation}
where $\ket{0}$ is the non interacting ground state and $\mathcal{N}$ is a normalization constant, the solution of the Hill-Wheeler equation for the $f$ amplitudes in the GCM ansatz will provide the exact wave functions. On the other hand, if one chooses to use as GCM coordinate the angle $\phi_2$ alone and fix $\phi_1$ to its HF value\footnote{From now, $\phi_1 = \phi_1^{(HF)}$, which is a constant given by the interaction parameter $\chi$.}, the GCM solution is not the exact one. This is representative of real situations where one does not know neither the right GCM coordinates nor their appropriate number. In the present case, one has to solve the Hill-Wheeler equation for $\phi_2$
\begin{equation*}
    \int d\phi_2' H(\phi_2,\phi_2')f(\phi_2') = E\int d\phi_2'N(\phi_2,\phi_2')f(\phi_2')
\end{equation*}
where\footnote{$\ket{\phi_2}$ is a shorthand notation for the state $\ket{\phi_1\phi_2}$ defined in \ref{eq:SD_3Lipkin}.}
\begin{equation*}
    H(\phi_2',\phi_2) = \bra{\phi_2'}H\ket{\phi_2}
\end{equation*}
and
\begin{equation*}
    N(\phi_2',\phi_2) = \bra{\phi_2'}\ket{\phi_2}
\end{equation*}
Using Eq. \ref{eq:SD_3Lipkin}, the overlap matrix will be given by
\begin{equation*}
    N(\phi_2',\phi_2) = \big(\vec{u}(\phi_2')\cdot \vec{u}(\phi_2)\big)^N
\end{equation*}
where $\vec{u}(\phi_2) = (\cos{\phi_2}\sin{\phi_1},\sin{\phi_2}\sin{\phi_1},\cos{\phi_1})\in\mathbb{R}^3$ is an unitary vector given in spherical coordinates. It can be shown that the overlap matrix depends only on the difference $\phi_2-\phi_2'$, so their eigenfunctions will be plane waves \cite{ring_schuck}
\begin{equation*}
    u_p(\phi_2) = \frac{1}{\sqrt{2\pi}}e^{-ip\phi_2}
\end{equation*}
and their eigenvalues will be given by
\begin{equation*}
    n_p = 2\pi\sum_{k=\abs{p},k+=2}^{N,N-1}\frac{1}{2^k}(\sin^2{\phi_1})^k(\cos^2{\phi_1})^{N-k}\binom{N}{k}\binom{k}{\frac{1}{2}(p+k)}
\end{equation*}
Using the fermion operators corresponding to the orbital basis which define the Slater determinants of Eq. \ref{eq:SD_3Lipkin}, we can compute the Hamiltonian matrix elements
\begin{equation*}
    H(\phi_2',\phi_2) = \epsilon N f(\phi,\phi')^{N-2} \bigg(f(\phi,\phi')g(\phi,\phi')-\frac{\chi}{2}h(\phi,\phi')\bigg)
\end{equation*}
where
\begin{align*}
    f(\phi,\phi') &=\vec{u}(\phi_2')\cdot \vec{u}(\phi_2) \\
    g(\phi,\phi') &= \sin^2{\phi_1}\sin{\phi_2'}\sin{\phi_2}-\cos^2{\phi_1'} \\
    h(\phi,\phi') &= 2\sin^2{\phi_1}\cos^2{\phi_1}\\ &+\sin^4{\phi_1}(\cos^2{\phi_2'}\sin^2{\phi_2}+\sin^2{\phi_2'}\cos^2{\phi_2}) \\
\end{align*}

Following \cite{ring_schuck}, if we compute the Hamiltonian elements in the basis given by the eigenfunctions $u_p(\phi_2)$ we obtain: 
\begin{widetext}
\begin{align*}
    &\bra{p'}H\ket{p} = \frac{\epsilon N}{\sqrt{n_{p'}n_p}}
    \bigg\{\sum_{0\leq k_1+k_2\leq N-1}\binom{N-1}{k_1,k_2}a^{k_1+k_2}c^{N-k_1-k_2-1}
    \bigg(a\mathcal{I}^{(k_1,k_2+1)^*}_{p'}\mathcal{I}^{(k_1,k_2+1)}_{p}-c\mathcal{I}^{(k_1,k_2)^*}_{p'}\mathcal{I}^{(k_1,k_2)}_{p}\bigg) \\
    &-\frac{\chi}{2}\sum_{0\leq k_1+k_2\leq N-2}\binom{N-2}{k_1,k_2}a^{k_1+k_2}c^{N-k_1-k_2-2}
    \bigg(r^2\mathcal{I}^{(k_1,k_2)^*}_{p'}\mathcal{I}^{(k_1,k_2)}_{p} +
    a^2\big(\mathcal{I}^{(k_1+2,k_2)^*}_{p'}\mathcal{I}^{(k_1,k_2+2)}_{p}+
    \mathcal{I}^{(k_1,k_2+2)^*}_{p'} \mathcal{I}^{(k_1+2,k_2)}_{p}\big)  \bigg)
    \bigg\}
\end{align*}
with
\begin{equation*}
    \mathcal{I}^{(k_1,k_2)}_{p} = \sqrt{2\pi}\frac{1}{2^{k_1}}\bigg(\frac{i}{2}\bigg)^{k_2}
    \sum_{q_1=-\sfrac{k_1}{2}}^{\sfrac{k_1}{2}}\sum_{q_2=-\sfrac{k_2}{2}}^{\sfrac{k_2}{2}}
    (-1)^{\sfrac{k_2}{2}-q_2}\binom{k_2}{q_2+\frac{k_2}{2}}\binom{k_1}{q_1+\frac{k_1}{2}}
    \delta_{2q_1+2q_2+p,0}
\end{equation*}
\end{widetext}

and
\begin{equation*}
    \binom{N}{k_1,k_2} \equiv \frac{N!}{k_1!k_2!(N-k_1-k_2)!}
\end{equation*}
Once all the above quantities have been evaluated, one ca solve the discrete diagonalization problem $\sum_p\bra{p'}H\ket{p}g_p=Eg_{p'}$ to obtain the weight function 
\begin{equation*}
    f(\phi_2) = \sum_{p,n_p\neq 0}g_p\frac{1}{\sqrt{n_p}}u_p(\phi_2)
\end{equation*}
and the final GCM state $\ket{\psi_{GCM}} = \int d\phi_2f(\phi_2)\ket{\phi_2}=\sum_{pq}C_{pq}^{(GCM)}\ket{pq}$ where
\begin{align*}
    C_{pq}^{(GCM)} = &\sqrt{\binom{N}{p,q}}(\sin{\phi_1})^{p+q}(\cos{\phi_1})^{N-p-q} \\ &\sum_{k,n_k\neq 0}\frac{g_k}{\sqrt{n_k}}\mathcal{I}^{(p,q)}_k
\end{align*}
The reduced four orbital state can be computed using Eqs. (\ref{eq:01_RDM}), (\ref{eq:02_RDM}) and (\ref{eq:12_RDM}) where $C_{pq}\rightarrow C_{pq}^{(GCM)}$.


\section{\label{sec:n_1__n_2__partition}Results for \{$n_1,n_2$\} subsystem}


In this appendix we discuss the results obtained for the \{$n_1,n_2$\} partition (illustrated in Fig. \ref{fig:3Lipkin_model_scheme}).

\begin{figure}[h]
\includegraphics[width=0.5\textwidth]{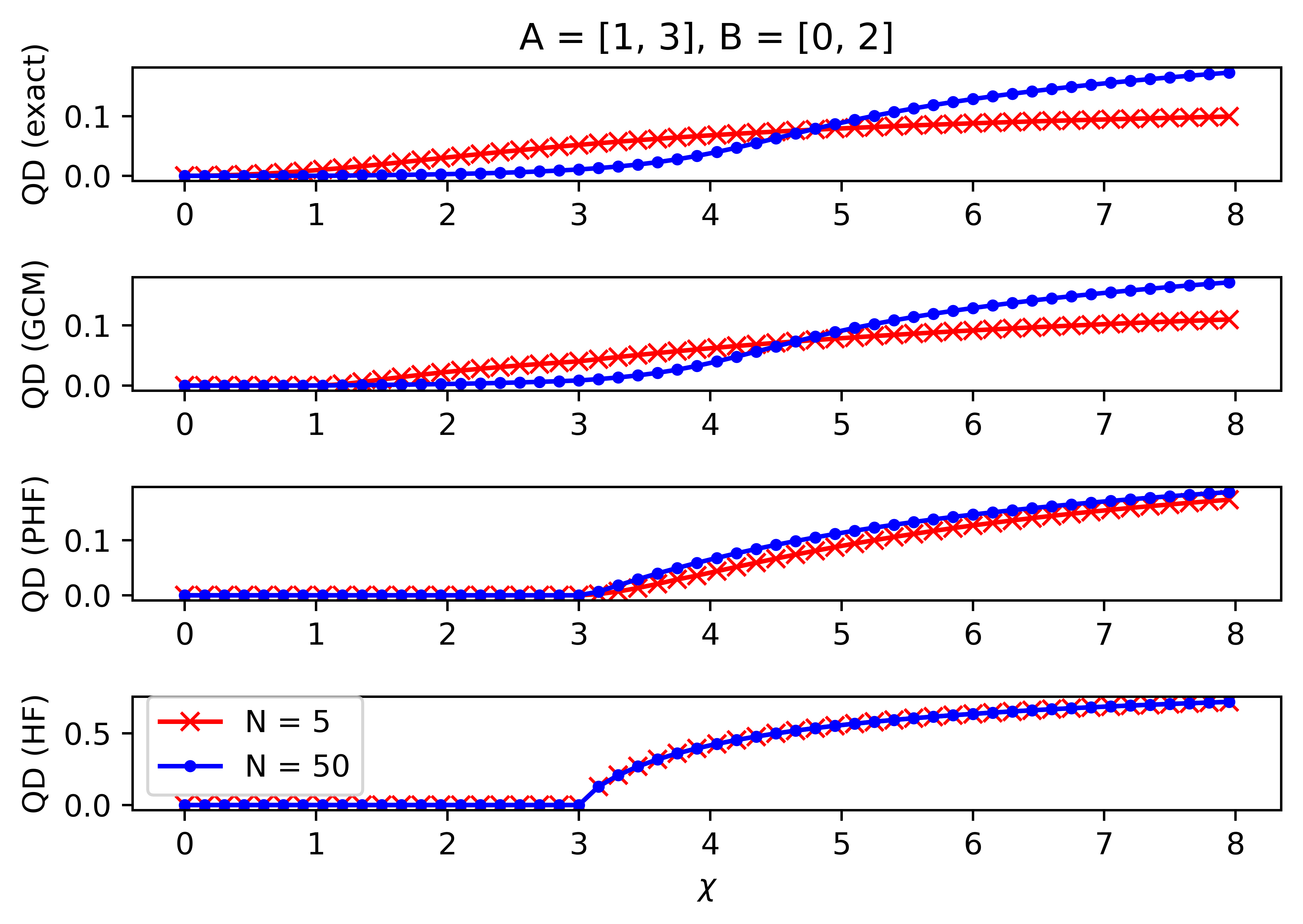}
\caption{Quantum discord for $A=\{1,3\}$ and $B=\{0,2\}$ in the exact (up), GCM (middle up), PHF (middle down) and HF (down) solutions.}
\label{fig:partition_12/A13_B02}
\end{figure}

\begin{figure}[h]
\includegraphics[width=0.5\textwidth]{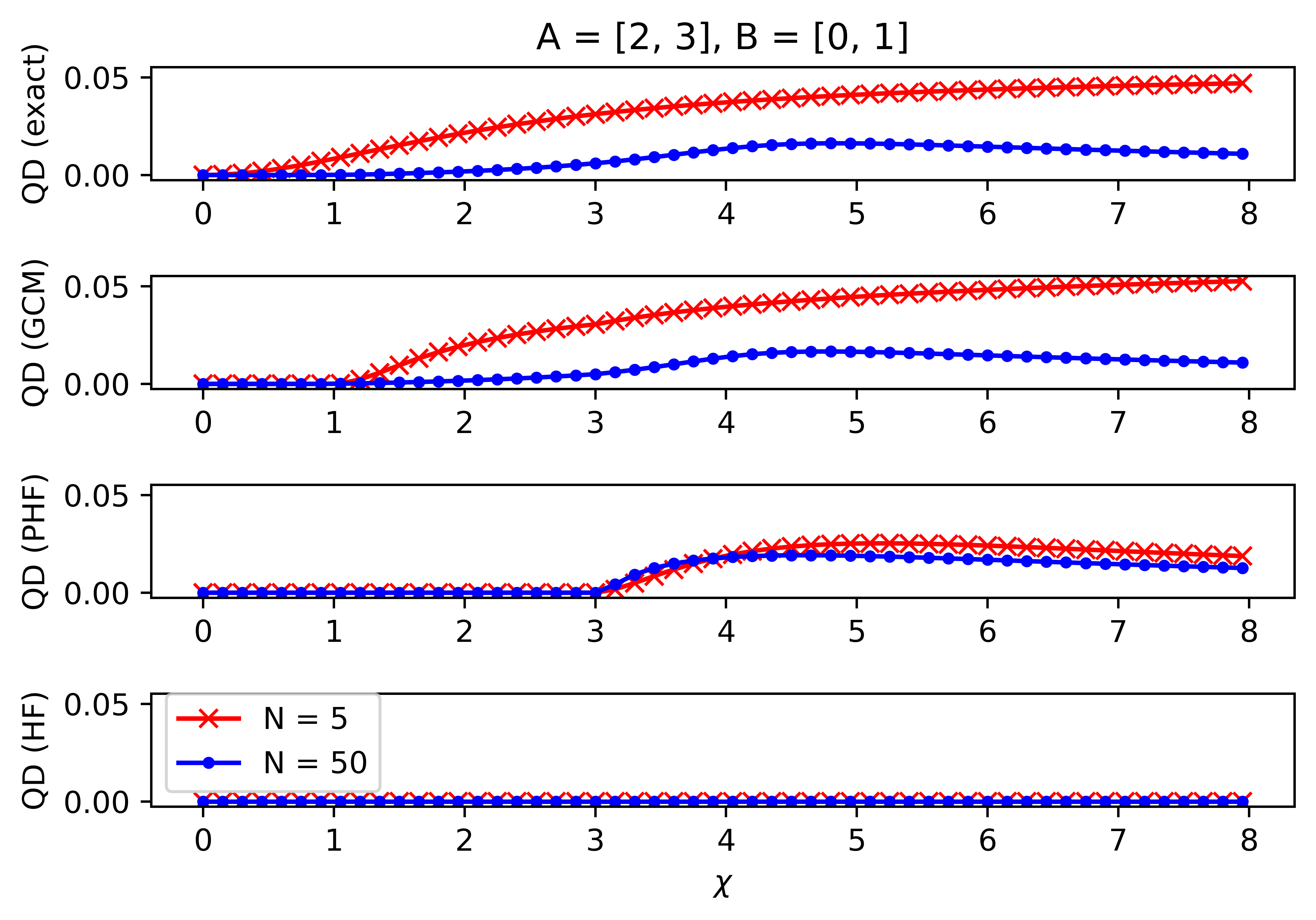}
\caption{Quantum discord for $A=\{2,3\}$ and $B=\{0,1\}$ in the exact (up), GCM (middle up), PHF (middle down) and HF (down) solutions.}
\label{fig:partition_12/A23_B01}
\end{figure}

In Fig. \ref{fig:partition_12/A13_B02}, the HF and PHF solutions do not present quantum correlations for $\chi<3$  since the second energy level is not populated in that case. The GCM solution is practically identical to the exact one because the QPT corresponding to the second energy level occurs at $\chi = 3$, and (specially if the particle number is large) the coordinate kept fixed to the HF value is related to the first energy level, whose QPT occurs at $\chi = 1$. 

One observes in Fig. \ref{fig:partition_12/A23_B01} that the HF solution does not catch correlations because the HF orbitals do not mix orbitals with different degeneration. Since the projection breaks the Slater determinant structure, the PHF is able to incorporate more quantum correlations. As in the previous case, and for the same reasons, the GCM solution is very accurate specially if $\chi>1$. 


\bibliography{bibliography} 
\end{document}